\documentclass[bibyear]{aa} 
\pdfoutput=1

\usepackage{graphicx}
\usepackage{txfonts}
\begin{document}
   \title{On the dust content of galaxy clusters}

   \subtitle{}

   \author{C. M. Guti\'errez\inst{1,2},
         M. L\'opez-Corredoira\inst{1,2}}
\institute{
$^1$ Instituto de Astrof\'\i sica de Canarias,
E-38205 La Laguna, Tenerife, Spain\\
$^2$ Departamento de Astrof\'\i sica, Universidad de La Laguna,
E-38206 La Laguna, Tenerife, Spain\\
}

\offprints{cgc@iac.es}
\titlerunning{On the dust of galaxy clusters}
\authorrunning{Guti\'errez \& L\'opez-Corredoira}

   \date{Received xxxx; accepted xxxx}

  \abstract
{} 
  {We present a study to estimate the dust content in galaxy clusters.}
  {This was done by using  one the most complete existing catalogues of galaxy clusters based on Sloan Digital Sky Survey (SDSS) data and following two methods: the first one  compares the colours of samples of galaxies in the background of clusters with those of galaxies in the field. Using this method, we have explored  clustercentric distances up to 6 Mpc. The galaxies used in this first method were selected from the SDSS-DR9, among those having reliable photometry and accurate estimation of photometric redshifts. The results are largely independent of the galactic cut applied. At $\mid b\mid > 20^\circ$, the sample contains 56\,985 clusters in the redshift range $0.05  < z<0.68$ (the mean redshift is 0.30) and $\sim 5.3\times 10^6$ galaxies.  The second  method computes the contribution of dust in clusters of galaxies to the far infrared sky. That is estimated indirectly by measuring the effect of clusters in the $E(B-V)$  extinction map.}
  {Using the first method, we did not find any dependence with clustercentric distance in the colours of background galaxies. As representative of the whole results, the surface integral of the excess of colour $g-i$ in three rings centred in the clusters and with radius  0-1, 0-2, and 0-3 Mpc  is $-3.7\pm 3.5$, $+3.2\pm 6.8$, and $-4.5\pm 10.1$ milimag Mpc$^2$, respectively. This allows us to constrain the mass of dust in the intracluster media, $M_{dust} < 8.4\times 10^9$ M$_\odot$ (95\% C. L.) within a cluster radius of 3 Mpc. With the second method, which averages the extinction of all clusters, we find a surface integral of the excess of colour $g-i$ of $3.4\pm 0.1$ millimag Mpc$^2$. From the extinction and redshift of each cluster, we obtain 0.13 Jy and $(1.46\pm 0.03)\times 10^{45}$ erg s$^{-1}$ for the mean flux and luminosity  at 100 $\mu$m. This  is $\sim 60$ times the far infrared luminosity of a Milky Way-like galaxy. By assumming similar properties for the dust, we can estimate a total dust mass per cluster of $\sim 2\times 10^9$ $M_\odot$, which is compatible with the hypothesis that the dust is within the spiral galaxies of a cluster. Separating the clusters in $5\times 5$ bins in redshift and richness, we confirm previous findings of a clear increase in luminosity with a redshift that agrees with the trend expected from current models.}
  {}


   \maketitle
%

\section{Introduction}

Intergalactic media in galaxy clusters does not offer a comfortable ambient for
dust grains to survive. Dust grains are progressively destroyed by the collision of particles 
(Draine \& Salpeter 1979). Although this process is relatively fast, the exact sputtering time  depends on the density and temperature of the media, as well as on the chemical composition and size of the
grains. For a wide range in temperature, the growth of dust by accretion is not enough to compensate the
sputtering, although several mechanisms like mergers, supernova winds, or ram-pressure stripping can
inject dust continuously into intergalactic regions of clusters. Little is known from an observational point of view;  several pioneer studies (Zwicky 1962; Karanchetsev \& Lipovetskii 1969; Bogart \&
Wagoner 1973; Boyle et al. 1988; Romani \& Maoz 1992) compared the attenuation of the light as a
function of wavelength in objects situated in the background of clusters with those of objects in the
field.  Using single cases or small  samples of clusters, extinctions within the range  $\sim 0.1-0.4$
mags were estimated. A huge advance was made by Maoz (1995), who recognized  the existence of selection
effects in some of these previous studies and put a constraint $<0.05$ mag on the excess of colour of
the light of  background galaxies crossing  foreground Abell clusters. Xilouris et al. (2006) discover a
systematic shift in the colour of background galaxies viewed through the intergalactic medium of the
nearby M81 group. This reddening coincides with atomic, neutral gas that is
previously detected between the
group members. Myers et al. (2003) find anticorrelation between Quasi Stellar Objects (QSOs) and clusters but little reddening.
They suggest gravitational lensing as a possible explanation of the
anticorrelation. In the last decade,
several groups have used large compilations of data that are mostly Sloan Digital Sky Survey (SDSS) based to put much tighter constraints.
Essentially, all of them used a common technique based on the attenuation and reddening of the light of
background objects (galaxies and/or QSOs) when passing through a cluster. This is quantified by
comparing the same  properties of objects in the field. Nollenberg et al. (2003) used nearby,
medium, and poor Automated Plate Measuring Machine (APM) galaxy clusters and derived a 99\% C.L. upper limit in the reddening due to dust in
crossing clusters $A_R=0.025$ mag on 1.3 Mpc scales. Chelouche et al. (2007) compared the photometric
and spectroscopic properties of quasars behind clusters with those in the field. By using the SDSS-DR5
spectroscopic quasar sample (Schneider et al. 2007) and the catalogue of SDSS clusters obtained by Koester
et al. (2007), they detected an excess of colour $E(B-V)\sim 10^{-3}$ mags on Mpc scales. Bovy et al.
(2008) used a sample of SDSS luminous, early-type galaxies and the SDSS cluster catalogue obtained by
Berlind et al. (2006) and obtained restrictions $E(B-V)<3\times 10^{-3}$ and $8\times 10^{-3}$ mags on
scales of 1-2  and $<1$ Mpc, respectively. Muller et al. (2008), using a catalogue of 485 clusters and
galaxies extracted from the Red-sequence Cluster Survey (Gladders \& Yee 2005) and a catalogue of 90,000
galaxies with photometric redshifts in the range 0.5-0.8, did not find evidence of colour differences
with clustercentric distance and put a severe restriction on the average visual extinction of $<A_v>=
0.004\pm 0.010$ mags within a clustercentric distance  $R_{200}$. McGee \& Balogh (2010) explored the
presence of dust on large scales by using 70,000 low redshift SDSS galaxy groups  and clusters.
They claim the  detection of dust out to a cluster-centric distance of 30 Mpc $h^{-1}$.  Muller et al.
(2009) summarized the results of those studies (see their Table 1).

Other authors had followed a more direct approach searching for the contribution of
intracluster dust emission to the far infrared sky maps.  Of course, the total
IR emission of the cluster contains the contribution from  dust within  galaxy members
and within the intracluster media. First claims of direct extended spatial detections (Wise et al. 1993) in the infrared using IRAS   were controversial  with  several authors (e.g.
Stickel et al. 2002) pointing out that part of these detections could be due to 
dust within galaxy members of the clusters, or to Galactic cirrus. The most
complete study following this method was conducted by Montier \& Giard (2005) and
Giard et al. (2008) combining IRAS data of more than 10,000 clusters. Those authors
obtained a clear statistical detection of emission in the bands at 12, 25, 60, and
100 $\mu$m. According to the estimation of IR emission due to the different galaxy
populations in cluster member galaxies,  Roncarelli et al. (2010) concluded that
most  (if not all) of the signal detected comes from the emission of dust in 
cluster members. The MIPS (Multiband Imaging Photometer for Spitzer) offered a
significant quantitative improvement that has allowed  studies of single clusters
(Bai et al. 2007 on Abell 2029, and Kitayama et al. 2009 on Coma cluster). These
last authors  put the following constraints on the emission of  dust within the
central 100 kpc: $\sim 5\times 10^{-3}$, $6\times 10^{-2}$, and $7\times 10^{-2}$
MJy s$^{-1}$ at  24, 70 and 160 $\mu$m respectively. Following Kitayama et al.,
this translates to an expected visual extinction $A_v<0.02$ mag and a surface
mass density of dust of $\Sigma _d < 1.4\times 10^3$ $M_\odot$ kpc$^{-2}$.

The study presented here combines and extends the two approaches outlined above. In both cases, we follow
a statistical approach averaging the contribution of a large sample of clusters and background galaxies
selected from the SDSS survey.  The paper is structured as follows: After this introduction, Section 2
describes the properties of the samples used, the restrictions applied to get the final subsamples, and
the methology;  Section 3 analyzes the main results, estimates the amount of dust per cluster, and
compares it with the results presented by other authors and with theoretical expectations;  conclusions
are presented in Section 4.

\section{Samples and methodology}

As outlined in the previous section, we follow two methods: in the first, we study the additional
reddening of galaxies in the background of clusters as compared with similar galaxies in the field; in
the second, we estimate the contribution  of clusters of galaxies to the optical extinction maps. The
basic ingredients of our study are a map of extinction and catalogues of clusters and galaxies,
respectively. The sample of galaxies was obtained from the SDSS DR9 photometric catalogue
(Ahn et al. 2012). Several potential cluster catalogues were considered. On the basis to
optimize completitude and sky and redshift coverage, the sample of clusters obtained by
Wen et al. (2012) was chosen using the SDSS-III survey. The spatial coverage of that survey is  $\sim 14,000$
square degrees and contains 132,684 clusters in the redshift range $0.05 \leq z \leq 0.8$. According to
those authors, the catalogue is more than 95 \% complete for clusters  with a mass of $M_{200} > 10^{14}$
$ M_\odot$ in the range $0.05 \leq z \leq 0.42$ and contains a false detection rate less than 6 \%. 
The catalogue presents photometric redshifts for all the clusters, whilst only $\sim 30$ \%  have
determination of redshift based on the spectroscopy of their brighter cluster galaxy (an updated version
of the catalogue with 52,683 spectroscopic redshifts was presented by Wen \& Han 2013). From these
cases, it was estimated for the photometric estimation of redshifts, a systematic offset  $< 0.004$ and
a standard deviation $<0.018$. To complete the studies to lower redshifts, we also use the sample of 1,059 Abell clusters with spectroscopic redshifts (see
http://heasarc.gsfc.nasa.gov/W3Browse/galaxy-catalog/abellzcat.html).

The map for extinction used was the  well known and widely used  map by Schlegel et al. (1998). This map
covers the full sky and presents an estimation of the $E(B-V)$ extinction obtained from the analysis of
COBE DIRBE and IRAS data.  The pixel size is 2.37 arcmin with a $FWHM=6.1$ arcmin. Although the inmense
majority of the signal in that map is due to the Galactic extinction, any contribution from
the dust of extragalactic objects obviously leaves its own imprint\footnote{Only the contribution from
extragalactic objects with flux $>1.2$ Jy had been removed by Schlegel et al.}. According to such
authors, the maximum level of such possible extragalactic contamination is below 0.01 mags and should be
nearly uniformly distributed. 

\subsection{Method 1}

For each of the galaxies from SDSS-DR9, we downloaded the relevant information for this study:
equatorial coordinates, magnitudes in the $g$, $r$, and $i$ filters, photometric redshifts, 
and the corresponding errors as estimated by SDSS pipelines\footnote{From the various estimations of magnitudes
computed by SDSS pipeline, we follow the prescription in SDSS web pages and used the magnitudes  called
$modelmag$. SDSS computes photometric redshifts according to several algorithms; we used the one
recorded in the table $Photoz$  (for details see SDSS web pages).}. The number of objects catalogued as
galaxies in that survey and having $clean$ photometry\footnote{See SDSS web pages for an explanation of
this concept.} is $\sim 1.34\times 10^8$. We selected those galaxies with photometric errors $<0.1$ mag 
and consider a cut of 22 mag in the three filters, $g$, $r$, and $i$, considered. This dramatically reduces
the sample to 31,182,824 objects. We also removed a few outliers ($\sim 0.07$ \% of the sample) with
extreme $g-r$ ($<-0.5$ or $>2.6$ mags) or $g-i$ ($<0.0$ or $>3.9$ mags) colours.

As it is usual in extragalactic studies, we ignore the region close to the Milky Way plane ($\mid b \mid
< 20$ degrees) where any possible effect of reddening by clusters would be largely masked by the heavy
Galactic extinction at low latitudes. The exact cut in latitude is a compromise to select those regions
with  low Galactic extinction, whilst at the same time maintaining a sample of clusters large enough. We
ran numerous tests changing the Galactic latitude cut within the range 20-50 degrees, and in addition,
we removed those clusters from the analysis situated in regions with  mean extinctions $ > 0.05$ mags as
measured in  circle with a radius of 6 Mpc projected at the cluster distance. Nevertheless, none of the
results presented here is critically dependent from those restrictions.

The following step was to determine the relative position of each galaxy with respect to the sample of
clusters. For clusters, we use a conservative  error in the photometric redshift of 0.02 (see Wen et al.
2012), whilst we use the corresponding errors listed in the SDSS catalogue for galaxies. 
For each cluster, we selected all galaxies that are projected at clustercentric distances up to 6 Mpc. This
distance is $> 2\times R_{200}$  for all the clusters in our sample. The uncertainty in the estimation
of such distance due to the uncertainty in the photometric redshift of the clusters are $\sim $ 30, 20,
and 8 $\%$ at redshifts 0.05, 0.1 and 0.2, respectively, so this could introduce some filtering in the case
of a radial dependence of colours with clustercentric distance.

A galaxy was considered to be in the background of a cluster when  $\Delta z \equiv
z_{gal}-z_{cluster}>3\sqrt{\sigma_{z_{gal}}^2-\sigma_{z_{cluster}}^2}$.  The relatively large error in
redshift, due to the photometric technique used for the determination of the redshifts in the galaxies
and clusters, makes it impossible to know their relative position in the space of redshift for most ($\sim
2/3$) of the galaxies projected along the line of sight of a given cluster.

\begin{figure}
\includegraphics[width=9cm]{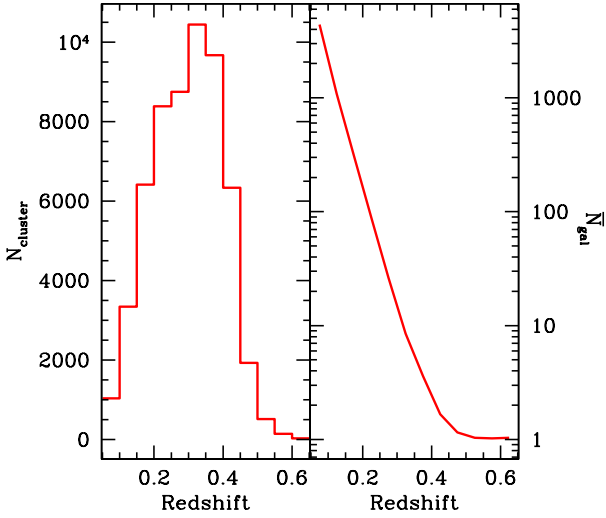}
\caption{$(Left:)$ Histogram of number of clusters as a function of redshift. $(Right:)$ Mean number of
galaxies in the background of clusters.\label{fig1}}
\end{figure}

Using a cut in galactic latitude of 20 degrees, the number of clusters and background galaxies selected
by this method are 56,985 and  5,287,825 (the corresponding numbers at $\mid b \mid >50$ degrees  are
36,473 and 3,520,658, respectively). Most of these galaxies behind a given cluster are projected in the
background of more than one cluster; they cross 2.1 clusters at projected
clustercentric distances  $<6$ Mpc on average. Figure \ref{fig1}  presents the redshift distribution of  clusters and
the mean number of background galaxies as a function of the redshift of the cluster. The cluster
distribution extends from redshift 0.05 to 0.6, having the maximum at $\sim 0.25$. This does not
exactly reproduce the distribution of redshifts in the Wen et al. (2012) catalogue, as clusters at very high redshift have not been considered because they do not have background objects  in the galaxy sample. As expected, the number of galaxies projected behind a given cluster depends strongly on the redshift of the cluster; this is basically a consequence of the cut in magnitude adopted for the galaxy sample.

\begin{figure}
\includegraphics[width=9cm]{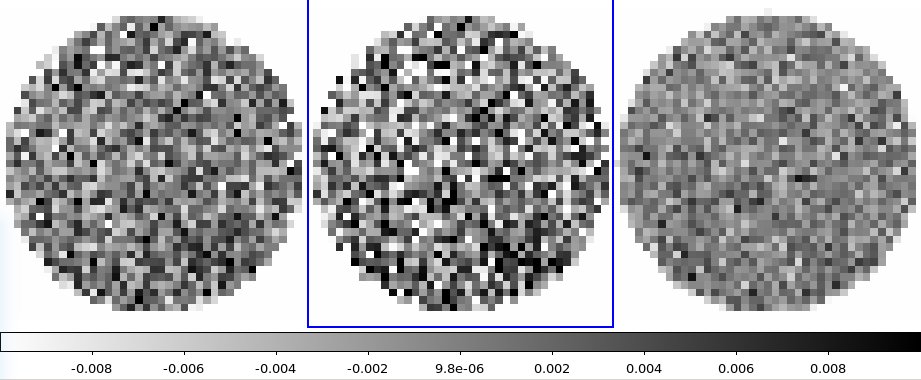}
\caption{Map of colours $g-r$, $g-i$, and $r-i$ (from left to right respectively) of
mean colours of galaxies as a function of the projected clustercentric distance. The
maps have a radius of 6 Mpc and a binning of 0.3 x 0.3 Mpc$^2$. The mean colour of each
map has been subtracted. The gray scale is linear and has an amplitude of 0.02 mags. \label{fig2}}
\end{figure}

\begin{figure}
\includegraphics[width=9cm]{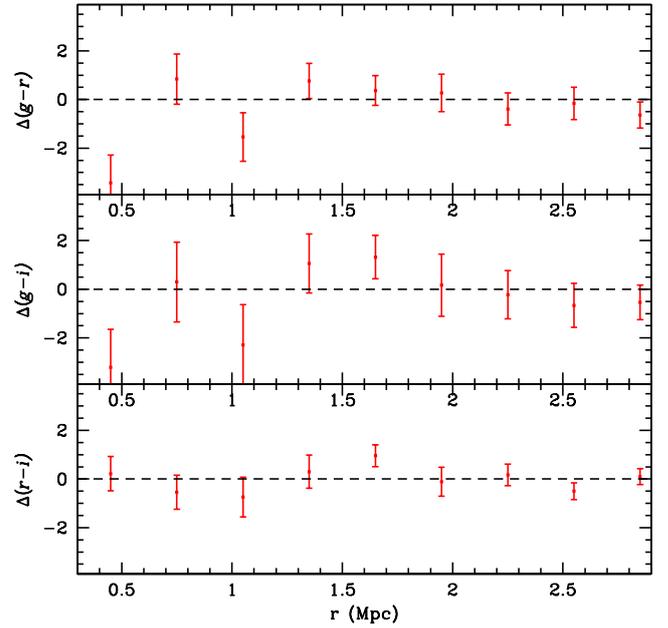}
\caption{Excess of colours as a function of projected clustercentric distance.
The units of the vertical axis are milimags.\label{fig3}}
\end{figure}

Maps of $(g-r)$, $(g-i)$ and $(r-i)$ colours and the corresponding radial profiles are shown
in Figs.~\ref{fig2} and \ref{fig3}, respectively. These maps represent the mean colours of galaxies in the background of clusters as a function of the projected clustercentric distance. The radius of the maps and the  binning are 6 Mpc and 0.3 Mpc x 0.3 Mpc, respectively. The mean
values of the colours are $1.348\pm 0.004$, $1.905\pm 0.006$, and $0.557\pm 0.003$ for $g-r$, $g-i$, and $r-i$, respectively. If we select only those objects at  $\mid b \mid >50$ degrees, the results are
similar with a slightly larger dispersion in each colour. None of the maps show any evidence of gradient
or structure. The limits on the excess of colours as a function of clustercentric distance were
estimated by taking a ring centred in each cluster as reference and having inner and outer radii of 4 and 5.5 Mpc, respectively. The error bars were computed from the dispersion of colours in a given bin and contain the uncertainty in the
estimation of the mean galactic extinction in addition to the uncertainty in the estimation of the mean colour. The results within  three  regions centred in the clusters 
are shown in Table~1. We do not detect excess of colours within any of the three radii considered. 
These results allow us to put 95\% C.L. upper limits  for the colour excess within a circle with radius 1
Mpc centred in the clusters of $2.69$, $3.37$, and $1.63$ milimags Mpc$^2$
in  $(g-r)$, $(g-i)$ and $(r-i)$, respectively.

\begin{table}
\begin{center}
\caption{Average excess of colours (in milimags Mpc$^2$) within different clustercentric radius.}
\begin{tabular}{cccc}
Rad (Mpc) & $\Delta (g-r)$ & $ \Delta (g-i)$ & $\Delta (r-i)$ \\
\hline
0-1 & $-2.23 \pm 2.46$ & $-3.67\pm  3.52$ & $-1.43 \pm 1.53$ \\
0-2 & $+1.54 \pm 4.73$ & $+3.20\pm  6.77$ & $+1.66 \pm 2.94$ \\
0-3 & $-4.73 \pm 7.06$ & $-4.52\pm 10.10$ & $+0.21 \pm 4.39$ \\
\end{tabular}
\end{center}
\end{table}

The above procedure introduces some degree of filtering in the  reddening by each cluster due
to the fact that a given galaxy could be in the background of more than one cluster, and then
its light  crosses  each foreground cluster at different clustercentric distances. The amount
of filtering would depend on the unknown spatial distribution of intracluster dust. To avoid
this uncertainty, several tests were conducted by modelling the dust spatial distribution. Here,
we only give the results of a very conservative approach that avoids any a priori dust
modelling by  considering only those galaxies lying in the background of just one cluster.
In doing that, the size of the sample drastically reduces to $\sim 7,000$ clusters and $1.8\times 10^5$ galaxies. We also do not detect any excess of signal within any clustercentric
distance. The corresponding 95\% C.L. limits are $120$, $39$, and $40$ milimags Mpc$^2$ for the
excess of colours $g-r$, $g-i$ and $r-i$ in a projected clustercentric radius of 1 Mpc,
respectively.

\subsection{Method 2}

We follow the method first proposed by Kelly \& Rieke (1990) and developed by Montier \& Giard (2005).
As discussed by these authors (an references therein), the detectability of individual clusters in IRAS
infrared maps is compromised by the low sensitivity of the maps, as compared to the signal expected
from clusters and to the noise confusion level of $\sim$ 1 MJy/sr due to extragalactic sources in those
maps. Instead of using the infrared maps as Montier \& Giard did, we used the extinction map $E (B-V)$
by Schlegel et al. (1998). Both approachs  are similar as the extinction $E(B-V)$ map is basically the
point-source subtracted IRAS 100 $\mu$m map, corrected to a reference temperature of 18.2 K using the
DIRBE temperature map, and multiplied by a constant $p = 0.0184\pm 0.0014$ mags /(MJy sr$^{-1})$. The estimations obtained using this method contain the contribution
of dust in the ICM and within galaxy members.

We built a $2\times 2$ deg (this corresponds to a radius $>3$ Mpc for all the clusters in the Wen et al.
sample) $E(B-V)$ map in galactic coordinates centred on each cluster, each rotated  by a random angle 
multiple of $\pi/2$, and averaged all the maps.  Given the large spatial density of clusters, a given
pixel in the extinction maps might contribute to several cluster maps. This introduces some degree of
correlation between pixels.  After many tests, we decided to limit the analysis to those clusters at a
galactic latitude $\mid  b \mid > 50$ degs. Although the mean galactic
extinction is 0.024 mags at $\mid b \mid >50$ degs, we  still flagged out all clusters situated in regions of particularly heavy
Galactic extinction, or those immersed in large gradients, i.e. those with a  mean level of extinction  $>0.05$ mags or a rms $>0.01$ mags within a square of 2 x 2 degrees centred in the cluster. None of the
results presented in the paper depends on the specific values of such constraints. The final sample
contains  52,667 clusters.

\begin{figure}
\includegraphics[width=9cm]{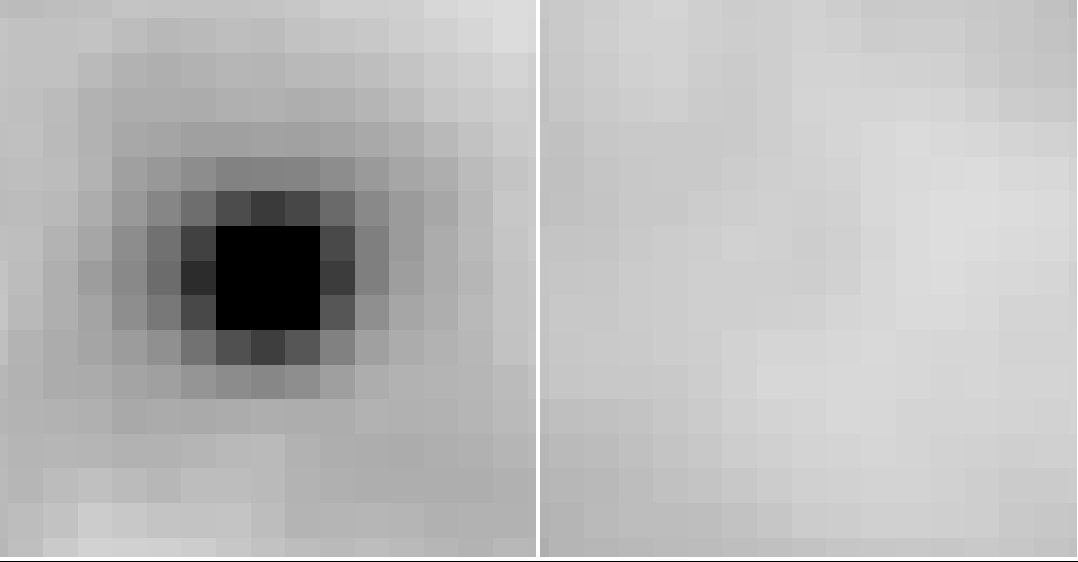}
\caption{($Left:$) Map of extinction (48 x 48 arcmin) obtained by averaging the maps centred
on each cluster. ($Right:$) A simulated map obtained shifting the position of each cluster
an angle randomly distributed in the range 1-2 degrees. \label{fig4}}
\end{figure}

\begin{figure}
\begin{center}
\includegraphics[width=6cm]{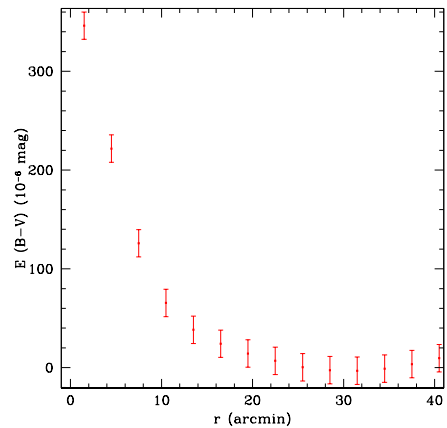}
\end{center}
\caption{Radial profile of the mean extinction in the stacked maps shown in previous figure. The
contribution of the Galactic extinction has been removed by averaging a ring of inner and outer radius of
24 and 40 arcmin respectively. \label{fig5}}
\end{figure}

Figure \ref{fig4} presents the average extinction map; this  shows a clear signal concentric with the position of the clusters. The radial profile of the extinction map is shown in Fig.~\ref{fig5}; this
indicates an extinction at the centre of the clusters   $E(B-V)=346 \times 10^{-6}$ mags. To assess the
reliability of the signal and the absence of significant systematics, we have created a new stacked map
by randomly and independently shifting the position of each cluster by an angle uniformly distributed in the range 1-2 degrees in galactic longitude. The resulting map is shown in Fig.~\ref{fig4} and
is compatible with noise random fluctuations.  From the uncertainties in the estimation of the
extinction from our own Galaxy and from the simulated results, we estimate a conservative uncertainty in
the central value of $\sim 50\times 10^{-3}$ mags; the main factor contributing to that is the absolute
level of the residual extinction from the Milky Way. The extinction  has a Gaussian profile width of
$FWHM=12$ arcmin.  Applying the conversion factor between $E(B-V)$ extinction and emission at 100 $\mu$m
quoted in Schlegel et al., the central emission translates into a mean  emission at 100 $\mu$m of
$(1.88\pm 0.08)\times 10^4$ Jy sr$^{-1}$. As it is demonstrated by separating the clusters in bins of
different richnesss and redshifts, this profile does not reflect the
spatial distribution of the cluster, but it is instead dominated by the instrumental resolution of the extinction map and  possibly additional filtering due to the method.

\section{Analysis}

\subsection{Method 1: Constraints on intracluster dust content}

For the case of a  Milky Way-like dust, in which
$A_V=1.73\Delta (g-i)_{\rm at\ rest\ in\ cluster}$ (Schlafly \& Finkbeiner 2011),
$A_\lambda \propto \lambda^{-1.4}$ approximately  in the interval from $g$
to $i$ filters. Hence, $\Delta (g-i)_{\rm at\ rest\ in\ cluster}
\approx \Delta (g-i)_{obs}(1+z_{\rm cluster})^{-1.4}$. 

Since
\begin{equation}
\int A_VdS=\int dV a_V
,\end{equation}\[
a_V(r)=1.086 \kappa _V \rho _{\rm dust}(r)
,\]\[
M_{\rm dust}=\int dV \rho _{\rm dust}(r)
,\]

the amount of dust within a given radius can be constrained using 
\begin{equation}
\frac{M_{dust}(r<R)}{M_\odot}=\frac{4.4\times 10^{15}}{\kappa_v} \int _0^R A_v dS 
,\end{equation}
where $\kappa_v=1.33\times 10^4$ g cm$^{-2}$ (Mathis 1990; Loeb \& Haiman 1997) and $S$
is the projected area in $Mpc^2$.  From the values quoted in Table~1 for the excess
$\Delta(g-i)$ within clustercentric distances of 1, 2 and 3 Mpc, we obtained the
corresponding $\int _0 ^R A_v dS$ values (see above) and estimated 95 \%  C.L. limits on the
mass of intracluster dust of $1.8\times 10^9$, $9.0\times 10^9$, and $8.4\times 10^9$
$M_\odot$ respectively. This corresponds to a limit on the projected density of dust of
$\Sigma <$ 573, 717 and 297 $\,M_\odot$ kpc$^{-2}$ within the clustercentric radius of 1, 2,
3 Mpc, respectively.

Assuming that our Galaxy has a dust exponential disc with scalelength 2.2 kpc (Drimmel \&  Spergel 2001), except in the inner 4 kpc to take the inner hole into account (L\'opez-Corredoira et al. 2004), and that
the extinction observed in the Galactic poles is $A_V=0.05$ mag (Schlegel et al. 1998), the Milky Way
would produce  $8.6\times 10^{-5}$ mag Mpc$^2$ observed from outside, which is 
around a dust mass of
$2.8\times 10^7$ M$_\odot $ (from Eq. 1). Davies et al. (1997), who used far-infrared emission, gets a similar number: $M_{dust}=3\times 10^7$ M$_\odot $. This means that we are constrained to have less than
$\sim 300$ times (95\% C.L.) the dust mass of the Milky Way in the intracluster medium. 

\subsection{Method 2: Infrared luminosities and total dust content}

The flux at rest is calculated as $F_{\nu,{\rm rest}}= (1+z)^{-1-\alpha}F_{\nu,{\rm observed}}$, given that the emission $I_\nu \propto \nu ^\alpha $; we take $\alpha =2$ (Schlegel et al. 1998). The extra factor
$(1+z)^{-1}$ stems from the increase of the frequency range, which gives an observed flux larger by a factor
$(1+z)$. 
The mean luminosity at 100 $\mu$m is $(1.46\pm 0.03)\times 10^{45}$ erg s$^{-1}$ for the
whole sample. The luminosity produced by the cold dust component of the Milky Way is $2.1\times
10^{43}$ erg s$^{-1}$ (Cox et al. 1986) or $2.6\times 10^{43}$ erg s$^{-1}$, according to
Davies et al. (1997). Assuming that the emission at 100 $\mu$m includes most of the dust
emission in clusters, the value quoted above would correspond to the emission of $\sim
60$  Milky Way galaxies  which is of the order of the number of spiral galaxies found
in clusters, which is $M_{dust}\sim 2\times 10^9$ $ M_\odot$ following the numbers of the
previous subsection.  

To compare the results of method 1 given in Table 1, we can convert the luminosity into the equivalent surface integral of the associated absorption by dust:
\begin{equation}
\int dS\,A_V=3.1\,\pi R^2 \overline{E(B-V)}
,\end{equation}\[
\overline{E(B-V)}=0.0184\, {\rm mag/(MJy/sr)}=\frac{F_{100,rest}}{\pi \alpha ^2},\,\alpha=\frac{R}{d_A}
,\]\[
\nu F_{100,rest}=\frac{L_{100,rest}}{4\pi d_L^2}
,\]
where $d_A$ is the angular distance, $d_L$ is the luminosity distance, which are
related by $d_A=\frac{d_L}{(1+z)^2}$. Hence,
\begin{equation}
\int dS\,A_V=1.58\times 10^{-47}\frac{L_{100,rest}({\rm erg/s})}{\langle (1+z)^4\rangle}
,\end{equation}
which gives us $\int dS\,A_V=(5.9\pm 0.1)$ millimag Mpc$^2$, or
$\int dS\,\Delta(g-i)=(3.4\pm 0.1)$ millimag Mpc$^2$ with the above number of $L_{100,rest}(erg/s)=(1.46\pm 0.03)\times 10^{45}$ and 
a value for our sample of clusters of $\langle (1+z)^4\rangle =3.87$. Using Eq. (2), we get again a dust mass of $2\times 10^9$ M$_\odot $.

According to the discussion by Roncarelli et al. (2010), the infrared emission at  100 $\mu$m
from galaxies in clusters is dominated by the emission of late-type spiral galaxies. Those
authors estimated a flux of $1904.8^{+617.1}_{-429.1}$ Jy in  a sample of 7476 MaxcBCG
clusters (Koester et al. 2007), or the value of
 $0.255^{+0.083}_{-0.057}$ Jy per cluster (we estimate
a mean flux per cluster of $0.130\pm 0.005$ Jy), and concluded that the values detected by
Giard et al. (2008) are roughly compatible with their estimations and leave little chance for
any component associated to intracluster dust. Our estimations agree with those by  Giard et
al. (see section 3.4) and then reinforces that scenario. The limit found by estimating the
possible reddening galaxies behind clusters (Method 1), although not very tight, are compatible
with this. The detection of intracluster dust using this method would require a larger
sample of background objects. 

\subsection{Abell clusters}

\begin{figure}
\includegraphics[width=9cm]{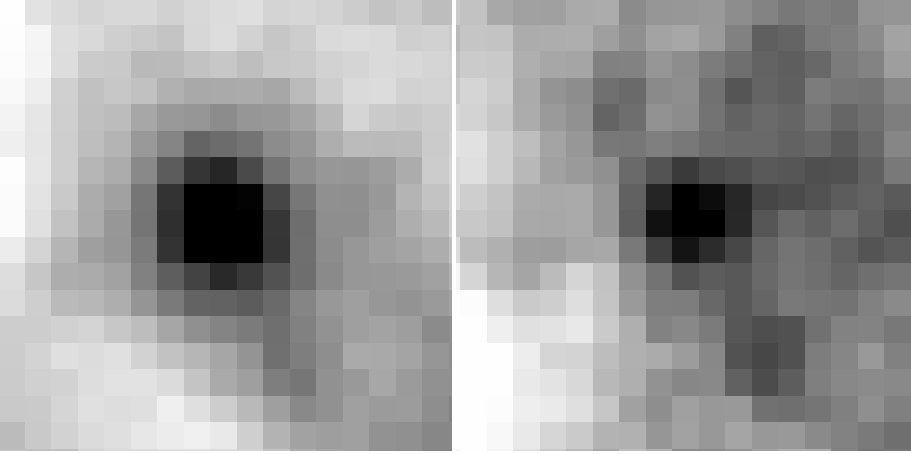}
\caption{($Left:$) Maps of extinction (48 x 48 arcmin) obtained by averaging the maps centred
on each cluster from the Abell sample. ($Right:$) The same restricted to those clusters at redshift
$z<0.05$.  \label{fig6}}
\end{figure}

As a test of the consistency of our results and to study the range in redshift $z<0.05$
that is uncovered by the Wen et al. (2012) sample, we do a similar analysis for the sample of
1,059 Abell clusters with measured redshifts (see
http://heasarc.gsfc.nasa.gov/W3Browse/galaxy-catalog/abellzcat.html). Considering the
whole sample and applying similar restrictions in galactic latitude and extinctions as
to the Wen et al. sample, we selected 485 clusters. The average extinction map
(Fig.~\ref{fig6}) centred
on each cluster shows a clear excess of signal with a central peak amplitude of
$E(B-V)=(1.16\pm 0.14)\times 10^{-3}$ mags that corresponds to a density of flux $\sim
(6.3\pm 0.8)\times 10^4$ Jy sr$^{-1}$. The radial profile has a $FWHM\sim 14$ arcmin,
which integrates up to a radius of
21 arcmin in which we obtain a 100
$\mu$m luminosity of $(0.79\pm 0.10)\times 10^{45}$ erg s$^{-1}$. If we restrict the
analysis to those clusters at redshift $z<0.05$, the number of clusters selected is
small (86 clusters). The stacked map (Fig.~\ref{fig6}) shows a central excess of extinction
$E(B-V)=(1.22\pm 0.35)\times 10^{-3}$ mags, and then a flux density and luminosity of
$(6.6\pm 1.9)\times 10^4$ Jy sr$^{-1}$ and $(0.12\pm 0.01)\times 10^{45}$ erg s$^{-1}$,
respectively. For the sample at $z<0.05$, the profile has a $FWHM\sim 15$ arcmin, and a
mean redshift 0.036. Assuming Gaussian profiles as a rough approximation, and that the
instrumental resolution is the one given by the profile obtained for the Wen et al.
clusters (i. e. $FWHM\sim 12$ arcmin, see next section), this gives a mean angular
extension of the dust emission in Abell clusters at redshifts $<0.05$ of $\sim 9$
arcmin, which corresponds to a radius of $\sim 0.6$ Mpc.

\subsection{Comparison with previous works}

\subsubsection{Method 1}

Comparison between our results and those found by other authors are not straightforward, due to the
different methods and samples used and the different spatial scales proved. In general, all the modern
works pointed out to a reddenig effect $<\, few\,\,10^{-3}$ mag of extinction per cluster.  Our method
relies on the hypothesis that the dust does not extend to distances larger than 6 Mpc (or at least that
it is not distributed uniformly on such scales). With all these cautions, our results are compatible
with those previous works by Nollenberg et al. (2003), Chelouche et al. 2007, Bovy et al. (2008), and
Muller et al. (2008), who present upper limits or marginal detections of excess at the level of
10$^{-3}$ mag. However, the work by  McGee \& Balogh (2010) explores the presence of dust on very large
scales by measuring colour excess of QSOs behind 70,000 low redshift SDSS galaxy groups  and clusters.
They claim the detection of dust out to a clustercentric distance of  30 Mpc $h^{-1}$. Although those
authors measured the excess of colours in very large scales (tens of Mpc), their results (see Fig. 4 in
their paper) indicate an excess of colour in $g-i\sim 5\times 10^{-3}$ mags, which is difficult to
reconcile with the radial profile of that excess, as it is shown in  Fig.~\ref{fig3}. We do not
have an explanation for that  inconsistency.

\subsubsection{Method 2}

As mentioned earlier, the only previous statistical work using a similar method was conducted by Montier \&
Giard (2005). These authors found an average central emission at 100 $  \mu$m of $(3.40\pm 0.14)\times
10^4$ Jy/sr, whilst we obtained  $(1.88\pm 0.08)\times 10^4$ Jy/sr. A direct comparison between both
amplitudes is not possible because we need to take into account the different angular width of both dust
profiles. The FWHM of their distribution at 100 $\mu$m has a $FWHM\sim 8$ arcmin (see their Fig.
5), whilst we obtained (see previous section) a $FWHM\sim 12$ arcmin. By assuming two dimensional Gaussian
profiles, the dilution factor, $(12/8)^2=2.25$, between both angular distributions and the ratio of the
peak amplitudes (1.8) translates into a flux density $\sim 25 \%$ in our case. This relative agreement is
quite remarkable considering the different samples of clusters used.  The reasons for our comparatively wider profile must come from the use of the Schlegel et al. maps with
a $FWHM\sim 6.1$ arcmin instead of the IRIS maps (Miville-Deschenes \& Lagache 2005) with a resolution
$FWHM\sim 4.3$ arcmin used by  Montier \& Giard. Additionally, some extra widening in our distribution
could be due to the several improvements of IRIS maps and in particular the better destripping.
This could also account for the reason that the relative  uncertainties in both studies are similar despite the fact of the higher number of clusters we have used.

\begin{figure}
\includegraphics[width=9cm]{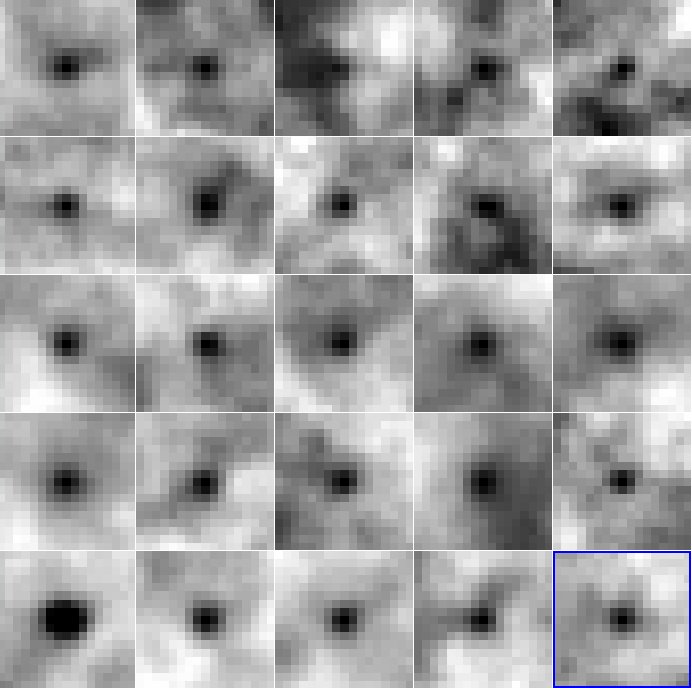}
\caption{Map of extinction (48 x 48 arcmin) obtained averaging the maps centred
on each cluster. Each map corresponds to a given range in richness and redshift.
Richness increases from top to bottom, while redshift increases from left to right. 
The mean values of richness and redshift in each bin are those quoted in Table~2.\label{fig7}} 
\end{figure}

\begin{figure}
\begin{center}
\includegraphics[width=6cm]{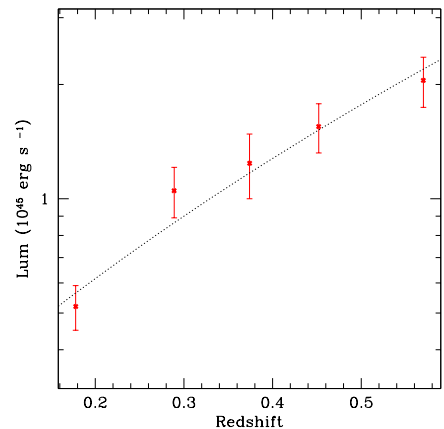}
\end{center}
\caption{Estimated mean cluster luminosity at 100 $\mu$m versus redshift. The dotted line corresponds to a
fit $L=L_o(1+z)^\alpha$ with $\alpha=4.73$\label{fig8}}
\end{figure}

\begin{table}
\begin{center}
\caption{Cluster Luminosities at 100 $\mu$m as a function of redshift and mass.}
\begin{tabular}{cccc}
$N_{clust}$ & Mass  ($10^{14}M_\odot$) & $z$ & L($10^{45}$ erg s$^{-1}$)  \\
\hline
   
1773	 &     0.633	  &	0.189	 & $  0.50 \pm       0.17$ \\ 
1772	 &     0.632	  &	0.297	 & $  0.59 \pm       0.20$ \\ 
1845	 &     0.632	  &	0.382	 & $  0.54 \pm       0.32$ \\ 
2047	 &     0.633	  &	0.462	 & $  1.16 \pm       0.43$ \\ 
1574	 &     0.634	  &	0.574	 & $  1.13 \pm       0.50$ \\ 
2234	 &     0.742	  &	0.186	 & $  0.33 \pm       0.11$ \\  
2081	 &     0.739	  &	0.297	 & $  0.94 \pm       0.34$ \\  
2276	 &     0.740	  &	0.381	 & $  0.99 \pm       0.38$ \\  
2635	 &     0.740	  &	0.463	 & $  0.93 \pm       0.37$ \\  
2290	 &     0.744	  &	0.576	 & $  1.69 \pm       0.56$ \\  
2321	 &     0.911	  &	0.184	 & $  0.48 \pm       0.17$ \\ 
2231	 &     0.913	  &	0.298	 & $  0.81 \pm       0.31$ \\ 
2337	 &     0.914	  &	0.381	 & $  1.38 \pm       0.53$ \\ 
2637	 &     0.913	  &	0.463	 & $  1.87 \pm       0.66$ \\ 
2690	 &     0.916	  &	0.583	 & $  2.33 \pm       0.80$ \\ 
2056	 &     1.153	  &	0.184	 & $  0.61 \pm       0.20$ \\ 
1736	 &     1.154	  &	0.297	 & $  1.11 \pm       0.39$ \\ 
1868	 &     1.152	  &	0.380	 & $  1.32 \pm       0.49$ \\ 
2012	 &     1.148	  &	0.463	 & $  2.11 \pm       0.75$ \\ 
2189	 &     1.150	  &	0.586	 & $  1.98 \pm       0.71$ \\ 
2712	 &     2.407	  &	0.180	 & $  0.75 \pm       0.25$ \\ 
1995	 &     2.249	  &	0.296	 & $  1.57 \pm       0.54$ \\ 
1852	 &     2.121	  &	0.379	 & $  2.17 \pm       0.73$ \\ 
1879	 &     1.994	  &	0.462	 & $  1.86 \pm       0.65$ \\ 
1615	 &     1.940	  &	0.584	 & $  2.91  \pm      1.00$ \\	
\end{tabular}
\end{center}
\end{table}

\begin{table}
\begin{center}
\caption{Cluster luminosities at 100 $\mu$m as a function of redshift.}
\begin{tabular}{cccc}
$N_{clust}$ & Mass  ($10^{14}M_\odot$)  & $z$ & L($10^{45}$ erg s$^{-1}$)  \\
\hline
 9944  &   1.246 &  		0.178	 &  	$0.52\pm 0.07	   $  \\
10297   &   1.146 &  		0.289	 &  	$1.05\pm 0.16	   $  \\
 9722   &   1.098 &  		0.374	 &  	$1.23\pm 0.24	   $  \\
10739   &   1.046 &  		0.452	 &  	$1.55\pm 0.23	   $   \\
11955   &   1.043 &  		0.570	 &  	$2.05\pm 0.31	   $   \\

\end{tabular}
\end{center}
\end{table}

\subsection{Dependence with richness and/or redshift}

To determine the origin and possible evolution of dust in clusters, it is interesting
to study the possible dependence of the dust emission with respect to mass and/or
redshift.  This was done by splitting the clusters in $5 \times 5$ bins. The limits
were chosen to have roughly the same number ($\sim 2\times 10^3$)  clusters
in each bin.  The masses, $M_{200}$, were obtained from the relation between richness and mass given by Wen et
al. (their equation 2). Although the resulting maps (Fig.~\ref{fig7}) are relatively noisy as compared
to the stacked map of the full sample presented in Fig.~\ref{fig4},  all the maps
show a clear detection of signal with the maximum at the centre of the maps. The main
problem having a low number of clusters is the existence of gradients or systematics
in the extinction map that prevents for a good estimation of the level of
Galactic extinction. The analysis of the beam profiles indicate that they are
dominated by the instrumental resolution and that the differences between them can be
adscribed to the effect of noise and in particular to the uncertainties in estimating the background level. After doing many tests, we conclude that the best
estimation of the luminosities in this case is obtained from the amplitude of the
central peak when a Gaussian profile is assummed. This width was obtained from the mean
value and rms  of the distribution of the 5 x 5 profiles. We obtain $FWHM=11.7\pm
1.9$ arcmin. The luminosities estimated according to the procedure outlined above are
quoted in Table~2. The horizontal and vertical axis represent redshift and mass,
respectively. The  five divisions in redshift correspond to central values (from left
to right) 0.19, 0.30, 0.38, 0.46, and 0.57, whilst the five vertical divisions
correspond to mean masses of (0.63, 0.74, 0.91, 1.15 and 2.2) $\times 10^{14}$
$M_\odot$, respectively.  There is a clear tendency to increase luminosity with
redshift and richness. This was quantified by fitting a function $L=L_0 \times
(1+z)^\alpha \times (M_{\rm cluster}/10^{14}M_\odot)^\beta$ erg s$^{-1}$.  The best fit
corresponds to values  $\log (L_0)=44.41\pm 0.10$,  $\alpha =4.7\pm 0.8$ and 
$\beta=0.64\pm 0.17$. The evolution found with redshift agrees  with the models by Le
Floch'h et al. (2005) ($\alpha=3.15\pm 1.6$) and Bay et al. (2007)
($\alpha=4.0^{+2.1}_{-2.2})$. Binning the maps in five bins in redshift and computing
the luminosities in a similar way, we obtained the results presented in Table~3.  As 
the five bins in redshift have roughly similar mean masses, we can ignore that
dependence and fit a function $L=L_0 \times (1+z)^\alpha$. The results are presented
in Table 3 and Fig.~\ref{fig8}.  The luminosity found for the Abell clusters (see 
section 3.3) at redshift $<0.05$ is $(0.12\pm 0.01)\times 10^{45}$ erg s$^{-1}$ (not
included in the plot) which follows the general tendency to decline luminosity at
lower redshifts. The ratio from the mean luminosity at redshift 0.57 (last bin in
Table~3) to the mean luminosty of the Abell clusters is a factor $\sim 17$. Assuming
that this ratio is entirely due to evolution in redshift (there could be some effect
due to differences in richness), this confirms the high evolution in redshift found by other authors using large samples
(e. g. Giard et al. 2008) or single clusters (Bay et al. 2007). These authors found a
ratio $\sim 17$ between the luminosities at 24$\mu$m of  Coma ($z=0.02$) and
MS1054-0321 ($z=0.83$). 

\section{Conclusions}

The main conclussions of our work are

\begin{itemize}

\item We have conducted a study to estimate the amount and distribution   of dust within galaxy
clusters. This has been done following two methods: ($i$)  analyzing the effect that such dust  produces
on the light of objects in their background, and ($ii$) analyzing the contribution of clusters to the
$E(B-V)$ extinction map by Schlegel et al. (1998).

\item We did not find evidence of additional reddening of background galaxies with respect to galaxies
in the field. Our analysis  imposes maximum limits in the excess of colour due to intracluster dust
extinction  $\int _0 ^R \Delta (g-i)dS=-3.67\pm 3.52$, $+3.20\pm 6.77$  
and $-4.53\pm 10.10$ milimags Mpc$^2$  within clustercentric distances of  1, 2 and 3 Mpc respectively.

\item Using the second method, we clearly detect the far infrared emission produced by
the clusters. The corresponding extinction profile can be characterized  by a Gaussian function with a
peak amplitude of $346\times 10^{-6}$ mags and a $FWHM\sim 12$ arcmin. The angular
profile is dominated by  instrumental effects due to the resolution of the extinction
map and the method used  and does not reflect the spatial distribution of the dust
within the clusters. Averaging the extinction of all clusters, we find a 
surface integral of the excess of colour $g-i$ of 3.4 millimag Mpc$^2$.

\item The above extinction corresponds to an average flux and luminosties at 100 $\mu$m
per cluster of 0.21 Jy  and $(1.46\pm 0.03)\times 10^{45}$ erg s$^{-1}$, respectively.
This signal can be explained as due to emission of $2\times 10^9$ $M_\odot$ of dust with
temperature of 20 K.

\item Our results do not allow us to exclude the existence of some intracluster dust,
but we constrain the maximum amount of dust to be a few tens the dust in Milky Way-like galaxies

\item Separating the clusters in $5\times 5$ bins in redshift and richness respectively, we found a clear
detection in each of them. Fitting a function $L=L_0 \times (1+z)^\alpha \times 
(M_{200}/10^{14}M_\odot)^\beta$ erg s$^{-1}$ erg s$^{-1}$.
The best fit corresponds to values  $\log (L_0)=44.41\pm 0.10$,  $\alpha =4.7\pm 0.8$, 
and  $\beta=0.64\pm 0.17$. The dependence in redshift agrees with previous studies.

\end{itemize}

Despite the results of previous studies and the work presented here,  new techniques exploring the dust
content in galaxy clusters for different subsamples of objects covering different ranges in mass  and
redshift are needed. Among the tools that can provide a deeper insight in the topic we would like to
mention those allowing a whole treatment of dust and gas and the use of the recent maps obtained by the
Planck/Herschel mission. As we have shown, the extinction due to dust in the intracluster media is too small
to be measured with current datasets of galaxies but can potentially be of
interest once new dataset are available.

\

\begin{acknowledgements}
Thanks are given to the anonymous referee for helpful comments. Thanks are given to C.-J. Lin (language editor of A\&A) for proof reading of the text.

Funding for SDSS-III has been provided by the Alfred P. Sloan Foundation, the Participating
Institutions, the National Science Foundation, and the U.S. Department of Energy Office of Science. The
SDSS-III web site is http://www.sdss3.org/. SDSS-III is managed by the Astrophysical Research Consortium
for the Participating Institutions of the SDSS-III Collaboration including the University of Arizona,
the Brazilian Participation Group, Brookhaven National Laboratory, Carnegie Mellon University,
University of Florida, the French Participation Group, the German Participation Group, Harvard
University, the Instituto de Astrofisica de Canarias, the Michigan State/Notre Dame/JINA Participation
Group, Johns Hopkins University, Lawrence Berkeley National Laboratory, Max Planck Institute for
Astrophysics, Max Planck Institute for Extraterrestrial Physics, New Mexico State University, New York
University, Ohio State University, Pennsylvania State University, University of Portsmouth, Princeton
University, the Spanish Participation Group, University of Tokyo, University of Utah, Vanderbilt
University, University of Virginia, University of Washington, and Yale University.  We have used also
the NED (NASA Extragalactic Database, http://nedwww.ipac.caltech.edu/)
\end{acknowledgements}

\end{document}